\documentclass[a4paper,11pt]{article}
\usepackage[lmargin=2cm,rmargin=2cm]{geometry}
\usepackage[utf8]{inputenc}
\usepackage{amsmath,amssymb,mathtools,amsthm,stmaryrd}
\usepackage{physics}
\usepackage{todonotes}

\usepackage{xspace}
\usepackage{qcircuit}
\usepackage{hyperref}
\usepackage{authblk}

\title{Comment to \emph{Spatial Search by Quantum Walk is Optimal for Almost all
		Graphs}}

\author{Ryszard Kukulski}
\author{Adam Glos\thanks{aglos@iitis.pl}}
\affil{Institute of Theoretical and Applied Informatics, Polish Academy of 
	Sciences, \\ Ba{\l}tycka 5, 44-100 Gliwice, Poland}
\date{}

\newcommand{\ii}{\mathrm i}

\newcommand{\Id}{\mathbb{I}}

\newcommand{\ER}{Erd\H{o}s-R\'enyi\xspace}
\begin{document}
	\maketitle 
	
	\begin{abstract}
This comment is to correct the proof of optimality of quantum spatial search for
\ER graphs presented in `Spatial Search by Quantum Walk is Optimal for Almost all Graphs' (\url{https://doi.org/10.1103/PhysRevLett.116.100501}). The authors claim that if
$p\geq \frac{\log^{3/2}(n)}{n}$, then the CTQW-based search is optimal for
almost all graphs. Below we point the issues found in the main paper, and
propose corrections, which in fact improve the result to $p=\omega(\log(n)/n)$ in case of transition rate $\gamma = 1/\lambda_1$. In the case of the proof for simplified transition rate $1/(np)$ we pointed a possible issue with applying perturbation theory.		
	\end{abstract}
	
In this comment we use the notation $f(n) = o(g(n))$ for $\lim_{n\to\infty}
f(n)/g(n) = 0$, and $f(n)=\omega(g(n))$ for $g(n)=o(f(n))$.
	
	\section{Convergence of $\lambda_1$} 
	
In the section III of Supplementary materials of \cite{chakraborty2016spatial}
the authors refereed to the Theorem 1 from \cite{furedi1981eigenvalues}  to
prove the convergence of $\lambda_1$. However, for $\gamma A$ the assumptions of
this theorem are not satisfied, i.e. the values $\mathbb{E}(\gamma A_{ij}) =
n^{-1}$, ${\rm Var}(\gamma A_{ij})=n^{-2}p^{-1}(1-p)$ are not constant. We
propose to use the Theorem 3 from \cite{chung2011spectra} which works for
$p=\omega(\log (n)/n)$ and which states that asymptotically almost surely
	\begin{gather}\label{eq:lambda1-convergence}
	|\lambda_1 - 1 + 1/n| \le 
	C\sqrt{\ln(n)/(np)},
	\end{gather}
for some constant $C$. This in turn gives  a.a.s. the required condition
$\lambda_1=1+o(1)$.

\section{Convergence of $\max_{i\geq 2}|\lambda_i|$ } 

In Supplementary materials, the Theorem 1.4 from reference \cite{vu2005spectral}
was applied to matrix $\gamma(A - \mathbb{E}(A))$ to show
	\begin{enumerate}
		\item $\| \gamma(A - \mathbb{E}(A)) \| = o(1)$ in section III,
		\item $\max_{i\geq 2}|\lambda_i| = o(1)$ in section IV.
	\end{enumerate}  
The authors applied this theorem for $p  \ge \frac{\ln^{3/2}(n)}{n}$, while the
Theorem 1.4 requires $p=\omega(\log ^2(n)/n)$. More precisely if we consider
matrix $\gamma(A - \mathbb{E}(A))$, then it is necessary to take (up to a
constant factor) $K = (np)^{-1}$ and it is optimal to choose $\sigma = \max
\left(\frac{1}{n \sqrt{p}}, \frac{\ln^2(n)}{n^{3/2} p}\right)$ up to the
constant factor. The first argument of $\max$ comes from the fact that $\sigma$
has to be an upper-bound on the variance of the matrix, and the second argument
comes from the assumption from the Theorem 1.4.  Taking the proposed $\sigma$,
we can show that $\| \gamma(A - \mathbb{E}(A)) \|  \leq \sigma \sqrt{n} +
\sqrt{K \sigma} n^{1/4} \ln(n) = o(1)$ only for $p=\omega(\log ^2(n)/n)$.
	
Let us show that the results are valid for $p=\omega(\log (n)/n)$. Studying the
proof of Theorem 1 from \cite{chung2011spectra} one can show that for
$\varepsilon \coloneqq \sqrt{\frac{\ln(n)}{np}} = o(1)$ we have $
P(\|\gamma(A-\mathbb{E}(A))\| \le \sqrt{\varepsilon}) \ge 1 - \varepsilon, $
which shows the point 1. above. For point 2., one can apply Theorem 3 from
\cite{chung2011spectra} to show that a.a.s. $\max_{i\geq 2}|\lambda_i| \le
\varepsilon = o(1)$.
	
	\section{Improper proof in Section IV of Supplementary materials}

In section IV the authors showed that for any vertex $\ket{w}$ the probability
of finding marked node converges a.a.s. to one. This was done by using
perturbation theory and approximating $\ket{s'}$ with $\ket{s}$ based on
$|\braket{s}{s'}|=1-o(1)$ which had been shown previously.

Such an approach leads to three contradictions. First, based on the
approximation of $\ket{s'}$ we can only have $\ketbra{s} =
(1-o(1))\ketbra{s'}{s'} +H'$ where $\|H'\| = o(1)$.  The $(1-o(1))$
approximation prohibits from replacing $\gamma_p$ with $1/(np)$ which has to be
done with precision $\order{\frac{1}{\sqrt n}}$, based on Eq.~(46) in
\cite{chakraborty2016spatial}. Second, $H'$ influences the perturbation error,
which is no longer $\order{\gamma_p^2\lambda_2^2}$, but only $o(1)$, which is is
irrelevant in the context of the main result of the paper. Third, working only
with approximation $\ket{s'} \approx \ket{s}$ and neglecting the value of
$\braket{w}{s'}$ could falsely improve the efficiency of finding nodes. To see
this, let $0<f(n)=o(n)$ and let $\ket{\varphi} = \frac{1}{\sqrt{n-f(n)}}
\sum_{i=0}^{n-f(n)-1} \ket{i}$. $\ket{\varphi}$ has $1-o(1)$ overlap with
$\ket{s}$ but nodes $w = n - f(n), \ldots, n-1 $ cannot be found with success
probability greater than $1/n$.

Below we propose a corrected proof.

\paragraph{Convergence $\braket{w}{s'}=1/\sqrt{n}(1+o(1))$.} There are two ways
to show the desired convergence. First, based on the Proposition 1 from
\cite{glos2018vertices}, a.a.s. all nodes satisfy this equality provided
$p=\omega(\log^3(n)/(n\log\log n))$. Another way is to prove that a.a.s. this
equality holds for \emph{almost} all nodes provided that $p =
\omega(\log(n)/n)$.

Following the proof of Lemma 2 from Supplementary Materials of
\cite{chakraborty2016spatial}, we can show that a.a.s. $\braket{s}{s'} \ge 1-2/n
- \varepsilon - \sqrt{\varepsilon} = 1 - o(1)$ provided that $p =
\omega(\log(n)/n)$. Based on this result we will show that for \emph{almost} all
vertices the equality $\braket{w}{s'} = \frac{1}{\sqrt{n}}(1+o(1))$ holds.
	
Let $\ket{s'} = \alpha \ket{s} + \beta \ket{s^\perp}$. Define
\begin{gather}
W(G) \coloneqq \{ w \in \{1,\dots,n\} 
\colon | \sqrt n \braket{w}{s'} - \alpha| \le \sqrt{\beta}\}.
\end{gather} 
For all $w \in W(G)$ we have
\begin{gather}
\begin{split}
|\sqrt{n}\braket{w}{s'} - 1| &\leq | \sqrt{n} \braket{w}{s'} - \alpha| + |1-\alpha| 
\leq \sqrt{\beta } +|1-\alpha | = o(1).
\end{split}
\end{gather}
Let $W^c(G) = \{1,\ldots,n\} \setminus W(G)$. We have
\begin{equation}
\beta^2 n = \sum_{w=0}^{n-1} |\beta \sqrt{n} \braket{w}{s'}|^2  \geq \sum_{w\in
	W^c(G)} |\beta \sqrt{n} \braket{w}{s'}|^2 = \sum_{w \in W^c(G)} |\sqrt n
\braket{w}{s'} - \alpha|^2 > \beta	|W^c(G)|,
\end{equation}
hence $|W^c(G)|< \beta n$, and therefore $|W(G)| \ge n(1- \beta) = n(1-o(1))$.
That means there are a.a.s. only $o(n)$ nodes which do not converge to
$\frac{1}{\sqrt{n}}$ sufficiently fast. 
	
\paragraph{The main result} 
	
Let $\lambda_1 = 1 + \nu$, where $|\nu| \leq \frac{1}{n}+ C\sqrt{\frac{\ln
		n}{np}}$. Then, we have
\begin{equation}
\begin{split}
|\bra{w} \exp(- \ii (- \ketbra{w}-\gamma A) t) \ket{s} |^2 &=  |\bra{w} 
\exp(-\ii (- \ketbra{w}-\gamma A) t) \ket{s'} |^2 - o(1) \\
&= |\bra{w} \exp(-\ii (- \ketbra{w}-\gamma A + \nu \Id) t) \ket{s'} |^2 - o(1)\\
&= |\bra{w} \exp(-\ii (- \ketbra{w}- \ketbra{s'} + \widetilde{H}) t) \ket{s'} |^2 - o(1),
\end{split}
\end{equation}
where $\widetilde{H}$ is orthogonal to $\ketbra{s'}$ and $\|\widetilde{H}\| \le
|\nu| + \max_{i\geq 2}|\lambda_i| \le 1/n + (1+C)\sqrt{\frac{\ln(n)}{np}} = o(1)
$. Using the perturbation theory we have
\begin{equation}
|\bra{w} \exp(- \ii (- \ketbra{w}-\gamma A) t) \ket{s} |^2 = |\bra{w} \exp(- \ii
(- \ketbra{w}- \ketbra{s'}) t) \ket{s'} |^2 - o(1),
\end{equation}
which effectively reduces the evolution to two-dimensional space. We will 
discuss the correctness of applying the perturbation theory in the next 
paragraph.
	
Below we follow a derivation similar to the one presented in section IV in
Supplementary materials in \cite{chakraborty2016spatial}. The effective
Hamiltonian takes the form $H_{\rm eff} \coloneqq -\ketbra{s'} - \ketbra{w}$.
From now, we will assume that $\varepsilon\coloneqq \braket{w}{s'} =
\frac{1}{\sqrt n}(1+o(1))$. Let $\ket{\bar s'} =
\frac{1}{\sqrt{1-\varepsilon^2}} \left(\ket{s'} -\varepsilon \ket{w}\right)$.
Then the Hamiltonian takes the form
\begin{equation}
H_{\rm eff} = \begin{bmatrix}
-1 +\order{1/n} & \frac{1}{\sqrt{n}}(1+o(1))\\
\frac{1}{\sqrt{n}}(1+o(1)) & -1  - \order{1/n}
\end{bmatrix}\approx   \begin{bmatrix}
-1 & \frac{1}{\sqrt{n}}\\
\frac{1}{\sqrt{n}} & -1 
\end{bmatrix},
\end{equation}
where we neglected all the terms of order $o(1/\sqrt n )$. By this, the success
probability can be approximated as $P_w(t) \approx \sin^2\left(\frac{t}{2\sqrt
	n}\right)$ \cite{chakraborty2016spatial}.

\paragraph{Comment on the application of perturbation theory} The above proof
follows the sketch presented in section IV in Supplementary Materials
\cite{chakraborty2016spatial}, where the authors utilized perturbation theory to
approximate the original evolution Hamiltonian with a simpler one. Below, we
present an example which suggests that such an approximation may be invalid.

Let $\gamma H + \ketbra{w}{w}$ be an evolution Hamiltonian with optimally chosen
$\gamma$ and $H$ having $1-o(1)$ spectral gap, and $\|H\|=1$. Then, it
can be represented as $\gamma \ketbra{\lambda_1}{\lambda_1} + \ketbra{w}{w} +
\gamma \tilde H$ with $\|\gamma \tilde H \| = o(1)$. Suppose $\tilde H =
\frac{1}{\log n} (\Id - \ketbra{\lambda_1})$. Note that such $\tilde H$
satisfies the requirements used in the section IV. However, $\Id$ has no effect
on the evolution and the Hamiltonian takes the form $\gamma(1+\frac{1}{\log n})
\ketbra{\lambda_1}{\lambda_1} + \ketbra{w}{w}$. However, $\gamma$ has to be
chosen within the precision $\order{1/\sqrt n}$, and thus the matrix $\tilde H$ affects the
precision of $\gamma$.

\begin{figure}\centering 
\includegraphics{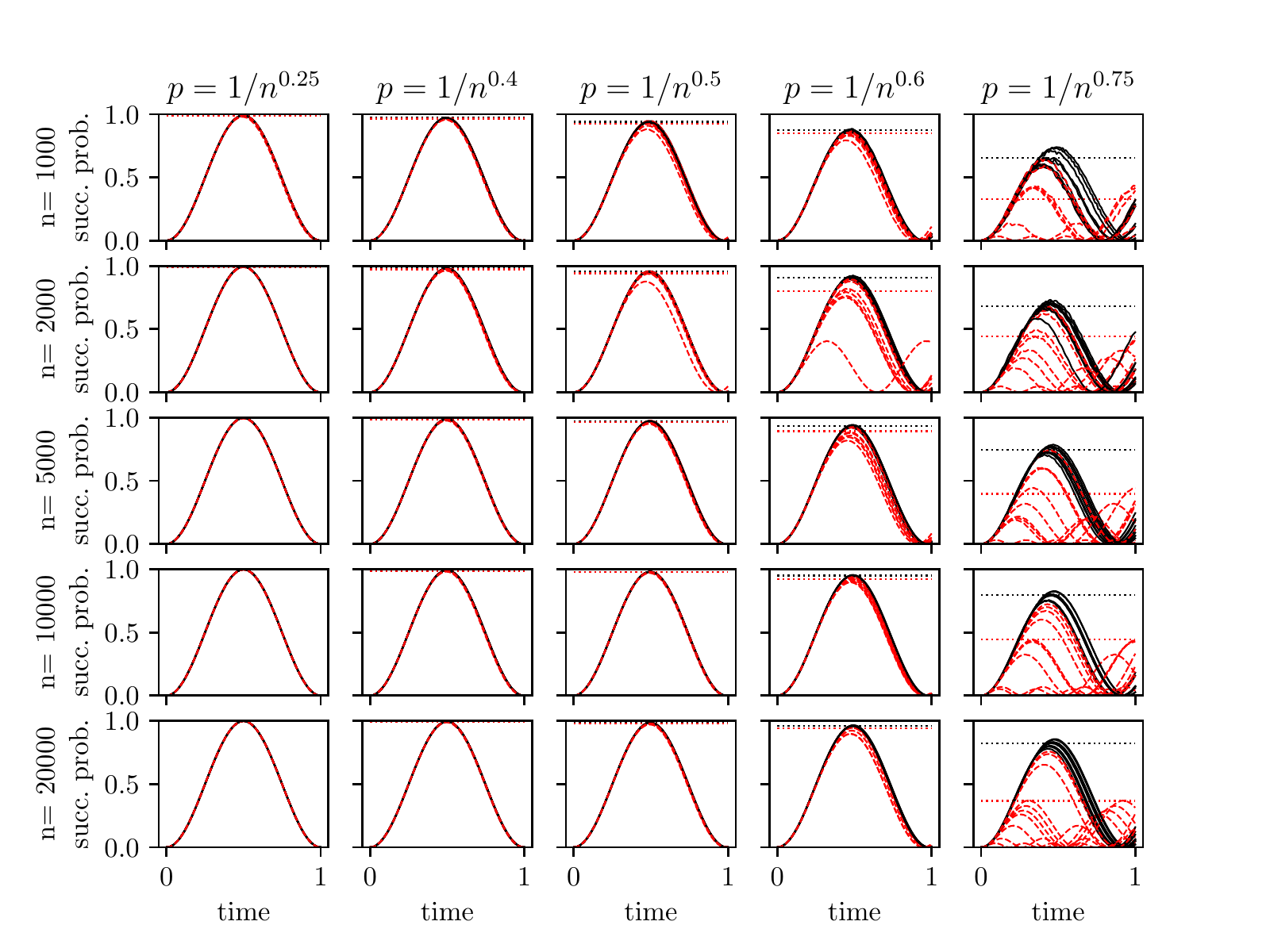}
	\caption{\label{fig:plot} A comparison between the success probabilities 
	for transition rate obtained from the proof in section I from Supplementary 
	materials (solid black lines) \cite{chakraborty2016spatial} and $1/(np)$ 
	(dashed red lines).  For each $(n,p)$ pair we took at random 10 \ER graphs. 
	Dotted lines are means of maximum success probability over graphs. The 
	time for each graph is rescaled to $[0,1]$, where $1$ corresponds to $2T$ 
	and $T$ is 
	the optimal time calculated according to the proof of Lemma 1. The code can be found on \url{https://doi.org/10.5281/zenodo.4055929}}	
\end{figure}

In Figure~\ref{fig:plot} we present a comparison of quantum search evolution between exactly derived
transition rate and $1/(np)$. We can see that for $p\geq 1/n^{0.6}$ the
approximation $1/(np)$ seems to be correct. It is less evident for $1/n^{0.75}$.

\section{Minor comments} 
	
Reference [14] in the main paper is not correct, since for normalized Laplacian
matrix $\mathcal {L}$ attains eigenvalue 2 for bipartite graphs. Hence the
proposed Hamiltonian $H_1 = \Id - \mathcal {L}$ has spectral gap equal to zero.
The constant normalized algebraic connectivity remains to be a sufficient
condition with a Hamiltonian $H_1 = \Id - \frac{2}{3}\mathcal {L} $.
	
Furthermore, in reference [14] the normalized algebraic connective is defined as
the second largest eigenvalue of normalized Laplacian, while it is defined as
the smallest positive eigenvalue (which for connected graph is the same as
second smallest eigenvalue).
	
\paragraph{Acknowledgments} AG was partially supported by National Science
Center under grant agreement 2019/32/T/ST6/00158 and 2019/33/B/ST6/02011. RK was
supported by the Polish National Science Centre under project number
2016/22/E/ST6/00062. The authors would like to thank Aleksandra Krawiec for
reviewing the manuscript.
	
	\bibliographystyle{ieeetr}
	\bibliography{comment_paper}

\begin{thebibliography}{1}

\bibitem{chakraborty2016spatial}
S.~Chakraborty, L.~Novo, A.~Ambainis, and Y.~Omar, ``Spatial search by quantum
  walk is optimal for almost all graphs,'' {\em Physical {R}eview {L}etters},
  vol.~116, no.~10, p.~100501, 2016.

\bibitem{furedi1981eigenvalues}
Z.~F{\"u}redi and J.~Koml{\'o}s, ``The eigenvalues of random symmetric
  matrices,'' {\em Combinatorica}, vol.~1, no.~3, pp.~233--241, 1981.

\bibitem{chung2011spectra}
F.~Chung and M.~Radcliffe, ``On the spectra of general random graphs,'' {\em
  The electronic journal of combinatorics}, pp.~P215--P215, 2011.

\bibitem{vu2005spectral}
V.~H. Vu, ``Spectral norm of random matrices,'' in {\em Proceedings of the
  thirty-seventh annual ACM symposium on Theory of computing}, pp.~423--430,
  2005.

\bibitem{glos2018vertices}
A.~Glos, A.~Krawiec, R.~Kukulski, and Z.~Pucha{\l}a, ``Vertices cannot be
  hidden from quantum spatial search for almost all random graphs,'' {\em
  Quantum Information Processing}, vol.~17, no.~4, p.~81, 2018.

\end{thebibliography}
\end{document}